\documentclass[11pt,twocolumn,twoside] {IEEEtran}

\usepackage{amsbsy}			%,color,multicol}
\usepackage{amsmath}
\usepackage{amssymb}
\usepackage{amsfonts}
\usepackage{booktabs}
\usepackage{multirow}
\usepackage{latexsym}
\usepackage{amsthm}
\usepackage{ifthen}
\usepackage{tabularx}
\usepackage{csquotes}
\usepackage{lipsum}
\usepackage{bm}
\usepackage{algorithm}	% For algorithms
\usepackage{algorithmic}
\usepackage{graphicx,epstopdf}
\usepackage{caption}
\usepackage{subcaption}
\usepackage[normalem]{ulem}
\usepackage{xspace}
\usepackage{paralist}		% compact and in-paragraph* lists
\usepackage{breqn}
\usepackage{cite}
\usepackage{color}
\usepackage[utf8]{inputenc}
\usepackage{textcomp}
\usepackage{arydshln}

\allowdisplaybreaks
\newtheorem{remark}{Remark}

\let\oldremark \remark
\renewcommand{\remark}{\oldremark\normalfont}
\newcommand{\bx}{{\bm x}}

\begin{document}

\bstctlcite{IEEEexample:BSTcontrol}

\title{Distributed Framework for Optimal Demand Distribution in Self-Balancing Microgrid}

\author{ Meenakshi Chatterjee  \\
              Department of Electrical Engineering \\
              Yale University, New Haven, CT 06511, USA \\
              Email: meenakshi.chatterjee@yale.edu
              % <-this % stops a space
%\thanks{} % <-this % stops a %space
}  % <-this % stops a %space

\maketitle

%%%%%%%%%%    Abstract         %%%%%%%%%%%%%%%%%%
%
\begin{abstract}
This study focusses on self-balancing microgrids to smartly utilize and prevent overdrawing of available power capacity of the grid. A distributed framework for automated distribution of optimal power demand is proposed, where all building in a microgrid dynamically and simultaneously adjusts their own power consumption to reach their individual optimal power demands while cooperatively striving to maintain the overall grid stable. Emphasis has been given to aspects of algorithm that yields lower time of convergence and is demonstrated through quantitative and qualitative analysis of simulation results. 
\end{abstract}

%%%%%%%%%%    Keywords        %%%%%%%%%%%%%%%%%%
\begin{keywords}
demand response, dynamic optimization, self-balancing, smart grids.
\end{keywords}

%\IEEEpeerreviewmaketitle

%%%%%%%%%%%%%%%%%%%%%%%%%%%%%%%%%%%%%%%%%%
%%%%%%%%%%    Section I    Introduction         %%%%%%%%%%%%%%%%%%
%%%%%%%%%%%%%%%%%%%%%%%%%%%%%%%%%%%%%%%%%%
%
\section{Introduction}
\label{sec:intro}

Generally, during peak time utility companies may request the buildings to cut down their power in exchange of some incentives. This would require building managers to first receive messages or phone calls from the utility company and respond with an appropriate strategy. To facilitate automation 
of the entire process, each building will need to quantitatively adjust its own strategy while cooperatively trying to make the grid more stable. In a self-balancing electrical grid, the total consumption of all the buildings in the grid should not exceed the total power consumption of the grid, i.e.,
\begin{align}
\label{eqn:1}
\sum_{i=1}^{n} P^c_i \leq P^{\text{max}}_G
\end{align} 
If the total demand exceeds $P^{\text{max}}_G$ all buildings must adjust their individual demand cooperatively such that the total remains below the grid maximum. Buildings will be able to consume power as per the final adjusted demand from the next time window. A building will not know what the other’s demand is. We assume a distributed model, where each building can communicate with its neighbors. Each building declares a price per unit time which it is willing to pay for the demand. Whoever is willing to pay more will get more. There is a utility function associated with each building, which is dependent on the willingness to pay parameter. Based on the price, each building tries to adjust their demand, so that their utility is maximized to reach their optimal demands under the constraint that the grid is maintained stable. In short, the goal should be to keep the total power consumption of all the buildings as close as possible to $P^{\text{max}}_G$ but strictly below it. 

In~\cite{kelly1998rate}, Kelly et al. have proposed a pricing scheme based on which a group of users shares the capacity of a network. In this scheme, end users are informed whether their packets are marked or not and accordingly they adjust their transmission rates. They have shown that this is Proportionally Fair Pricing scheme and can decentralize the global optimum allocation of congestible resources. Ganesh et al. proposed a modification of this in~\cite{ganesh2001congestion} wherein the switch assigns each packet a price instead of the mark. The mechanism of congestion prices has been used to provide both feedback and incentives to the end-systems. A market-based mechanism has been proposed in~\cite{paschalidis2011market} that allows the Smart Micro-Grid Operator (SMO) to control the behavior of internal loads through price signals and to provide feedback to the Independent System Operator (ISO). Reference~\cite{weng2011managing} proposed a method for building mangers to efficiently account for energy consumption and manage plug-loads in enterprise buildings during demand response events.

The rest of the paper is organized as follows. Section~\ref{sec:model} presents the models for static and dynamic adjustments of power demand. In Section~\ref{sec:simulation}, we numerically evaluate the performance of the models that we proposed in the previous section. Finally we conclude the paper in Section~\ref{sec:conclusion}.  

%
%%%%%%%%%%%%%%%%%%%%%%%%%%%%%%%%%%
%%%    Section II    Model of Optimal Demand Distribution         %%%%%%
%%%%%%%%%%%%%%%%%%%%%%%%%%%%%%%%%%%
\section{Model of Optimal Demand Distribution}
\label{sec:model}

We follow the principles of demand response model proposed by Fan et al. in~\cite{fan2011distributed}. Consider a discrete time slot system with time window T. Let the demand of the  building in time slot be. The constraint is that at convergence when optimal demands are reached, then 
\begin{align}
\label{eqn:2}
\sum_{i=1}^{n} x_i(t) < P^{\text{max}}_G
\end{align}
The unit price of demand at any time slot  is a function of aggregate demand, i.e.
\begin{align}
\label{eqn:3}
p(t) = f\left(\sum_{i=1}^{n} x_i(t)\right)
\end{align}
Based on the price information, each building will update its demand for the next time slot. A price function can be represented as~\cite{ganesh2001congestion},
\begin{align}
\label{eqn:4}
f(x) = a \left( \frac{x}{\mathcal{C}} \right)^k
\end{align}
where~$\mathcal{C}$ is the capacity of the network. The constant denotes the basic price of power and  controls how the price is influenced by the variation of aggregate demand with respect to the capacity. Each building has utility~$u_i \left( x_i(t) \right)$ associated with it in time slot. A typical utility function is chosen as the logarithmic one~\cite{kelly1998rate},
\begin{align}
\label{eqn:5}
u_i(x) = w_i \log(x)
\end{align}
where~$w_i$ is the willingness to pay (WTP) parameter for the~$i^{\text{th}}$ building. A building~$i$ will choose its demand to maximize~\cite{ganesh2001congestion},
\begin{align}
\label{eqn:6}
u_i \left( x_i(t) \right) - x_i(t) p(t)
\end{align}
Each building will adjust its demand for the next time slot as per the equation given by~\cite{fan2011distributed},
\begin{align}
\label{eqn:7}
x_i(t+1) = x_i(t) + \alpha \left( w_i - x_i(t) p(t)\right)
\end{align}
where is a parameter controlling the convergence rate.  is the price per unit given by,
\begin{align}
p(t) &= a \frac{\sum_{i=1}^{n} x_i(t)}{P^{\text{max}}_G} + s\left( \sum_{i=1}^{n} x_i(t) - P^{\text{max}}_G \right) \nonumber\\
\label{eqn:8}
& \qquad \qquad \qquad \qquad \times  u\left( \sum_{i=1}^{n} x_i(t) - P^{\text{max}}_G \right)
\end{align}
where~$s(.)$ is a sigmoid and~$u(.)$ is a unit step function. In the above pricing scheme, depending on the willingness to pay, a building is free to choose how it should respond to price information which is motivated by the assumption that each is trying to maximize equation~\ref{eqn:6} and the constraint that overall demand is less than~$P^{\text{max}}_G$. The sigmoid function in the price has been included to incorporate the extra steepness of an extra amount of penalty in the price. The unit step function ensures that this extra penalty is applied only when the demand exceeds the grid maximum. For a building the convergence of equation~\ref{eqn:7} to optimal demand implies that equation~\ref{eqn:6} has been maximized.
%
%%%%%%%%%%%    Sub-section  Static     %%%%%%%%%%%
\subsection{Static Adjustment of Power Demand}
\label{subsec:static}
The solution to the problem consists of the following parts: 

\begin{itemize}
	\item[$1.$] Each building need to know the total demand at any time slot so that it can estimate the price and hence adjust its demand under the constraint. The buildings will compute the global average of the demand using distributed averaging. The distributed averaging is done using the Best Constant~\cite{xiao2004fast} method. From the global average, it now can determine the total demand at that time.
	\item[$2.$] After knowing the total demand, each building will compute the estimate of the price. Based on the price information and its willingness to pay, it will adjust its demand at the next time slot as per equation~\ref{eqn:7}, where~$p(t)$ is given by equation~\ref{eqn:8}.
	\item[$3.$] Go to step 1, until equilibrium is attained.
\end{itemize}

The system will reach equilibrium when the demand of each building will converge to individual optimal demand subject to the constraint of the problem. The optimal demand of each building depends on its initial demand, price information and its willingness to pay. This optimal demand will then be distributed to each building which it will consume.

%
%%%%%%%%%%%    Sub-section  Dynamic     %%%%%%%%%%%
\subsection{Dynamic Adjustment of Power Demand}
\label{subsec:dynamic}
In the above approach, each time all buildings come to a consensus to know the total demand, adjusts it, and then again checks the total demand. This continues until the system reaches equilibrium, every building gets its optimal demand and the grid stays stable. Thus, the overall time taken to reach the optimal demand and the system to come to equilibrium may be high. An approach to reduce the overall time would be to allow each building to simultaneously estimate the global average of the demand and adjust their individual demand accordingly in one-time slot. In the next time slot, each building will update its own estimate as the weighted summation of the estimate received from its neighbors and the change in its demand~\cite{rajagopalan2011distributed, nedic2009distributed}. 
This approach can be summarized in the following steps:
\begin{itemize}
	\item[$1.$] Allow each building a time slot~$t$ to communicate its current demand~$x_i(t)$ to its neighbor and form an estimate of the global average~$x^{\text{est}}_i(t)$ at that time slot.
	\item[$2.$] Based on the estimated global average of each building, it estimates the price per unit demand at that time slot as in equation~\ref{eqn:8}. Now based on the price and WTP parameter, each building will update its demand following equation~\ref{eqn:7}. Let the change in demand for building~$i$ be denoted by~$\Delta_i$ where,
	\begin{align}
	\label{eqn:9}
	\Delta_i(t) = x_i(t+1) - x_i(t)
	\end{align}
	\item[$3.$] At the next time slot~$t=t+1$, the demand of some of the buildings have changed as the initial measurements that we have started with has changed for the next iteration. So, each building updates its own demand estimate as a weighted sum of demand estimate it receives from its neighbours and the change in its value. This is expressed as 	
	\begin{align}
	\label{eqn:10}
	x^{\text{est}}_i(t+1) = \sum_{j} w_{ij}x^{\text{est}}_i(t) + \Delta_i(t)
	\end{align}
	Equivalently, in vector notation, 
	\begin{align}
	\label{eqn:11}
	\bx^{\text{est}} (t+1) = W \bx^{\text{est}}(t) + {\bm{\Delta}}_i(t)
	\end{align}
	where~$W$ is the weight matrix obtained by the “Best Constant” Algorithm~\cite{xiao2004fast}.
	\item[$4.$] Go to step 1 until equilibrium is attained. 
\end{itemize}

%
%%%%%%%%%%%%%%%%%%%%%%%%%%%%%%%%%%
%%%    Section III    Simulation        %%%%%%
%%%%%%%%%%%%%%%%%%%%%%%%%%%%%%%%%%%
\section{Simulation Results}
\label{sec:simulation}
The above model has been simulated with a network of~$10$ buildings. The weight matrix has been generated using the best constant method, where the optimal constant is found out to be~$0.3135$. In the demand updating equation, $\alpha=0.05$. In the price equation, $a = 1$ and $k = 4$. For a time window, the initial demand for the~$10$ buildings has been chosen from a uniform distribution fro~$50$ to~$100$ as:
\begin{align*}
\bx(1) = \begin{array}{cccccc}
\bigl[57.3 & 98.1 & 75.2 & 85.7 & 90.9 \\
& 93.4 & 52.2 & 69.9 & 62.9 & 80 \bigr]
\end{array}
\end{align*}
Sum of the initial demand is~$765.9$. Corresponding~$P_G^{\text{max}}= 700$. The willingness to pay parameter~$\bm{w}$ has been randomly chosen as 
\begin{align*}
\bm{w} = \begin{array}{cccccccccc}
\bigl[45 & 98 & 67 & 80 & 90 & 93 & 50 & 50 & 57 & 72 \bigr]
\end{array}
\end{align*}
The following argument guides the above choice of~$\bm{w}$: the value of~$a$ is taken as 1, so we expect that in equilibrium, the price will also settle for a value close to but less than 1 as per equation~\ref{eqn:8}, because the total demand at that time would be less than 700. Let this optimal price be~$p_{\text{opt}}$. Maximizing equation~\ref{eqn:6} would then give the optimal demand,
\begin{align}
\label{eqn:12}
x_{i, \text{opt}} = \frac{w_i}{p_{\text{opt}}}
\end{align}
As in our case~$p_{\text{opt}} \approx 1$, the optimal demand should almost be equal to the WTP parameter~$w_i$. Since the total initial demand is well above the grid maximum, all buildings will need to adjust their individual demands, so that when the system finally converges to equilibrium, each will reach its individual optimal demand and at the same time the total should be close to but less than 700.

%
%%%%%%%%%%%    Sub-section  Static     %%%%%%%%%%%
\subsection{Static Adjustment of Power Demand}
\label{subsec:sim_static}
\begin{figure}[!t]
	\centering
	\includegraphics[width=0.5\textwidth]{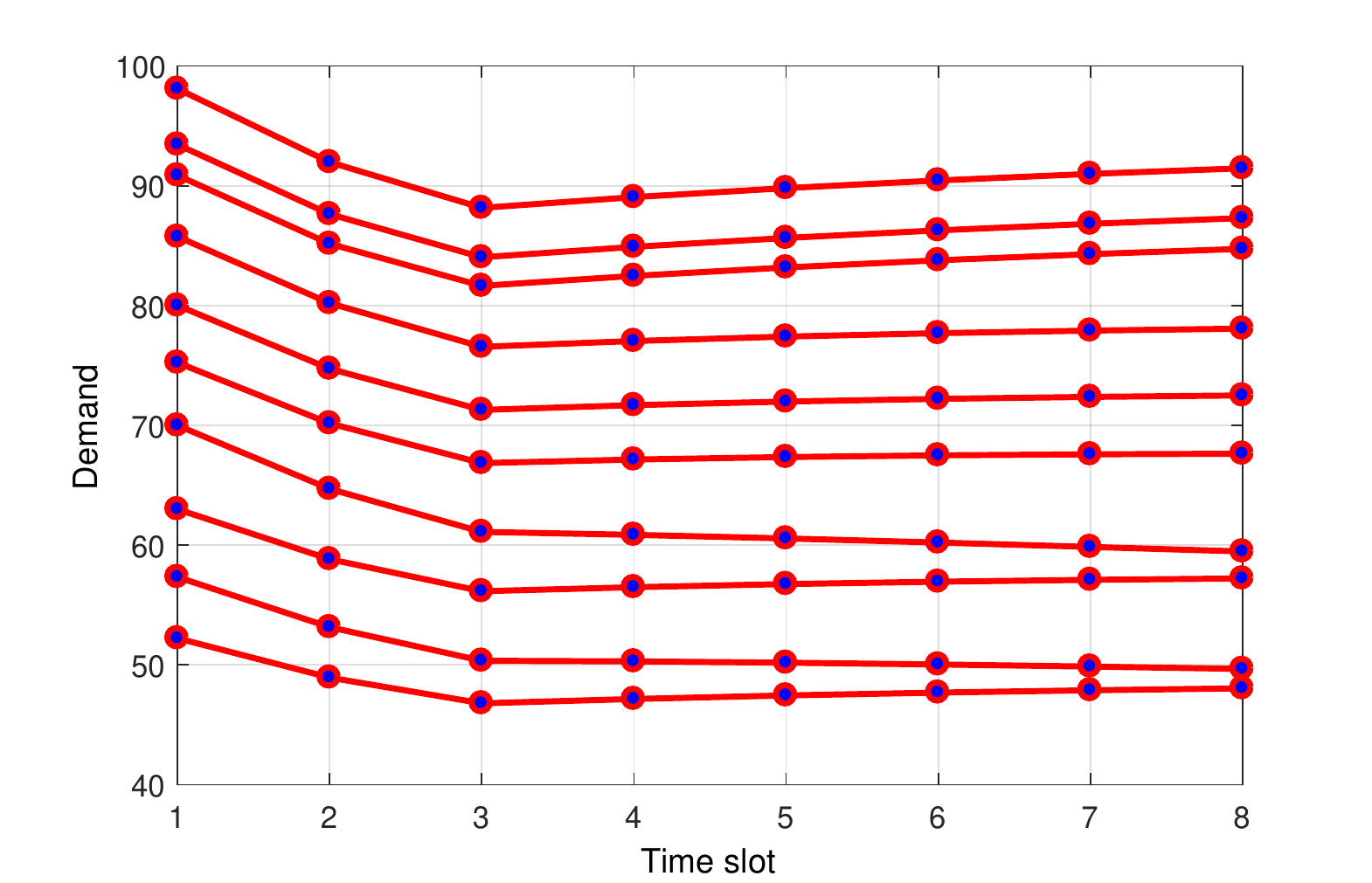}
	\caption{Demand adaptation of 10 buildings with time.}
	\label{fig:1}
	%	\hrule
\end{figure}
\begin{figure}[!t]
	\centering
	\includegraphics[width=0.5\textwidth]{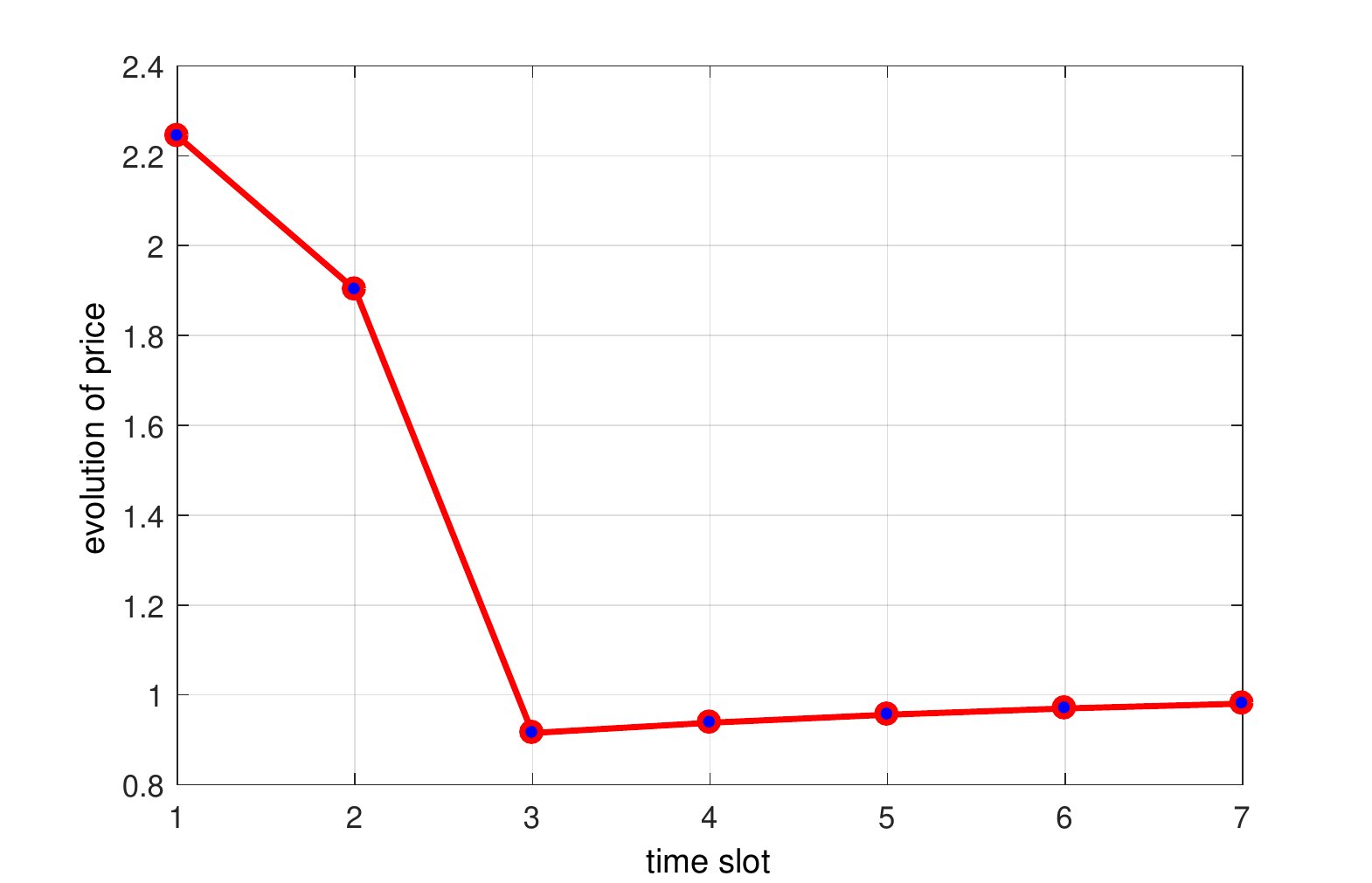}
	\caption{Evolution of price with time.}
	\label{fig:2}
	%	\hrule
\end{figure}
\begin{figure}[!t]
	\centering
	\includegraphics[width=0.5\textwidth]{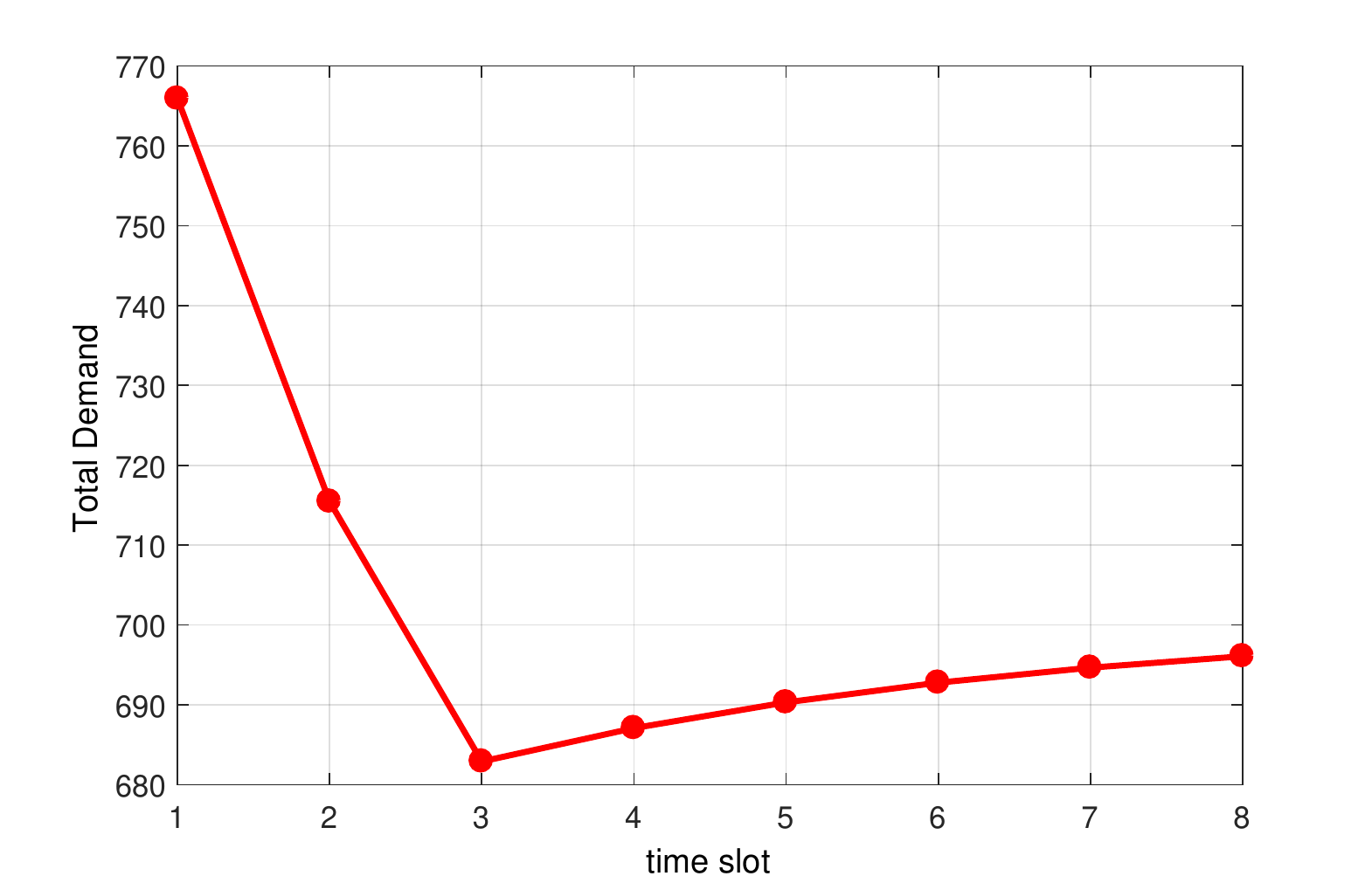}
	\caption{Variation of total demand with time.}
	\label{fig:3}
	%	\hrule
\end{figure}
\begin{figure}[!t]
	\centering
	\includegraphics[width=0.5\textwidth]{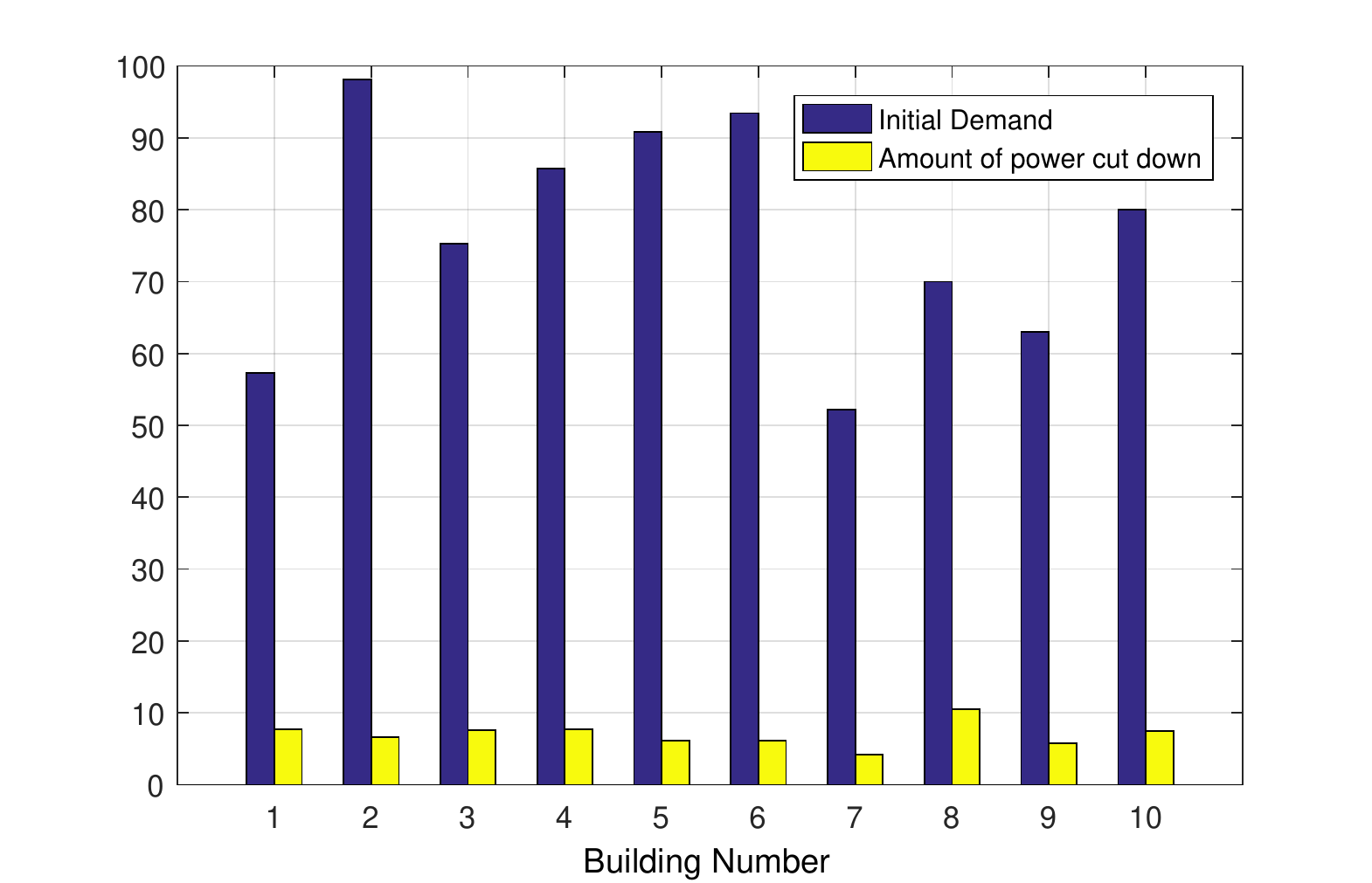}
	\caption{Demand cut down by the 10 buildings.}
	\label{fig:4}
	%	\hrule
\end{figure}

Fig.~\ref{fig:1} And Fig.~\ref{fig:2} shows demand adaptation for each building and the evolution of price per unit demand as a function of time slots respectively. Fig.~\ref{fig:3} shows the variation of the total demand with time slot. Initially, the overall demand which is equal to~$765.9$, is higher than~$P_G^{\text{max}} = 700$. So, the price is also high, with penalty being imposed depending on how much the total demand exceeded the maximum. Now, each building will adjust to lower its demand in the next time slot (time slot 2), after which the total demand in this slot, is nearly equal to 715 but above 700. So, the price is still high, but the penalty is less than what it was in time slot 1, as the amount by which the total demand exceeded 700 in this case is less than the previous case. All buildings will again adjust their demand for time slot 3, and check if the total demand is above 700 or not. In slot 3, the total demand falls below 700, and the price goes below 1 (as per equation~\ref{eqn:8}). So now that the grid is stable, some buildings might want to increase their demand depending on their willingness to pay parameter. The total demand would again start to increase and the overall price would then change accordingly. In this way, demand of each building will converge to a stable value, and the system will reach equilibrium. In slot 7 when the system is in equilibrium, the price is slightly below 1 and the total demand is slightly below the maximum of 700. Final values of the price and the total demand, when the system reached equilibrium, are 0.9804 and 696.0809 respectively. Thus, the demand is kept as close as possible to 700 at equilibrium, while not exceeding it and hence avoiding under-utilization of the grid capacity. The optimal demand is then finally available for consumption. Fig.~\ref{fig:4} shows the demand cut down by each building to reach the optimal subjected to the constraint of the problem.
%
%
%%%%%%%%%%%    Sub-section  Dynamic     %%%%%%%%%%%
\subsection{Dynamic Adjustment of Power Demand}
\label{subsec:sim_dynamic}
\begin{figure}[!t]
	\centering
	\includegraphics[width=0.5\textwidth]{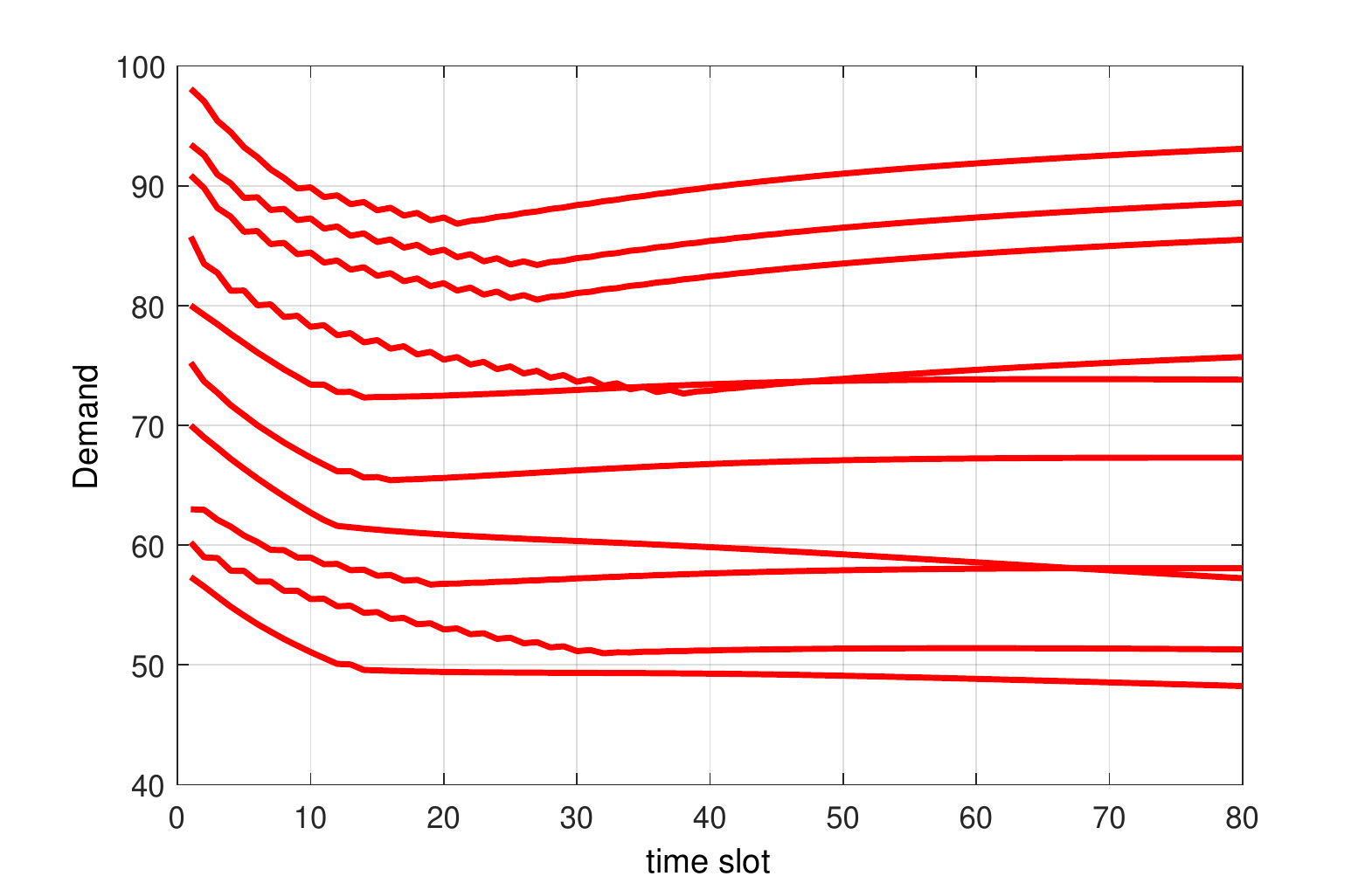}
	\caption{Demand adaptation of 10 buildings with time.}
	\label{fig:5}
	%	\hrule
\end{figure}

\begin{figure}[!t]
	\centering
	\includegraphics[width=0.43\textwidth]{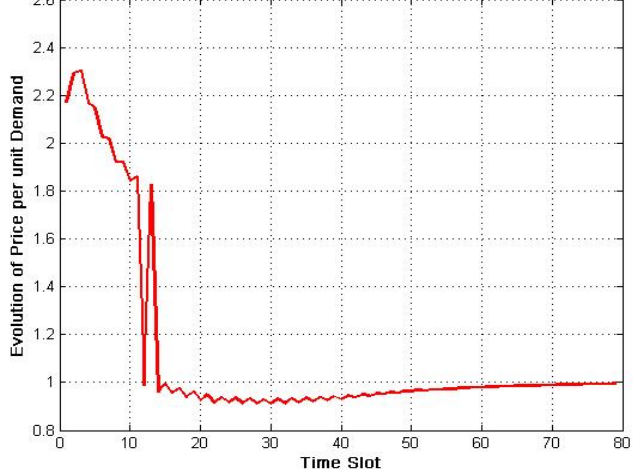}
	\caption{Evolution of price with time.}
	\label{fig:6}
	%	\hrule
\end{figure}

\begin{figure}[!t]
	\centering
	\includegraphics[width=0.43\textwidth]{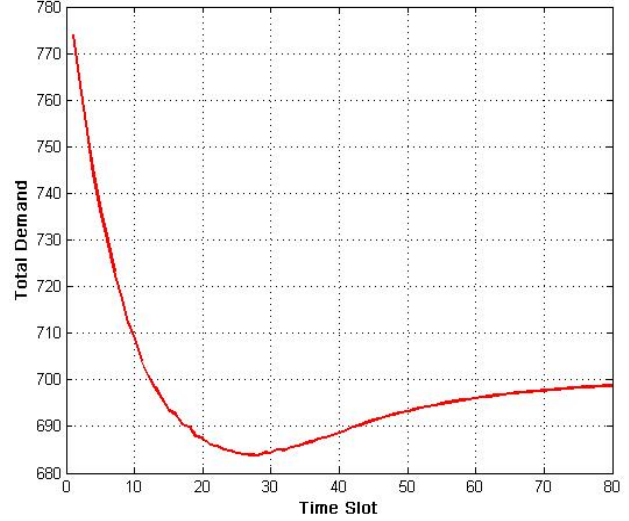}
	\caption{Variation of total demand with time.}
	\label{fig:7}
	%	\hrule
\end{figure}

\begin{figure}[!t]
	\centering
	\includegraphics[width=0.5\textwidth]{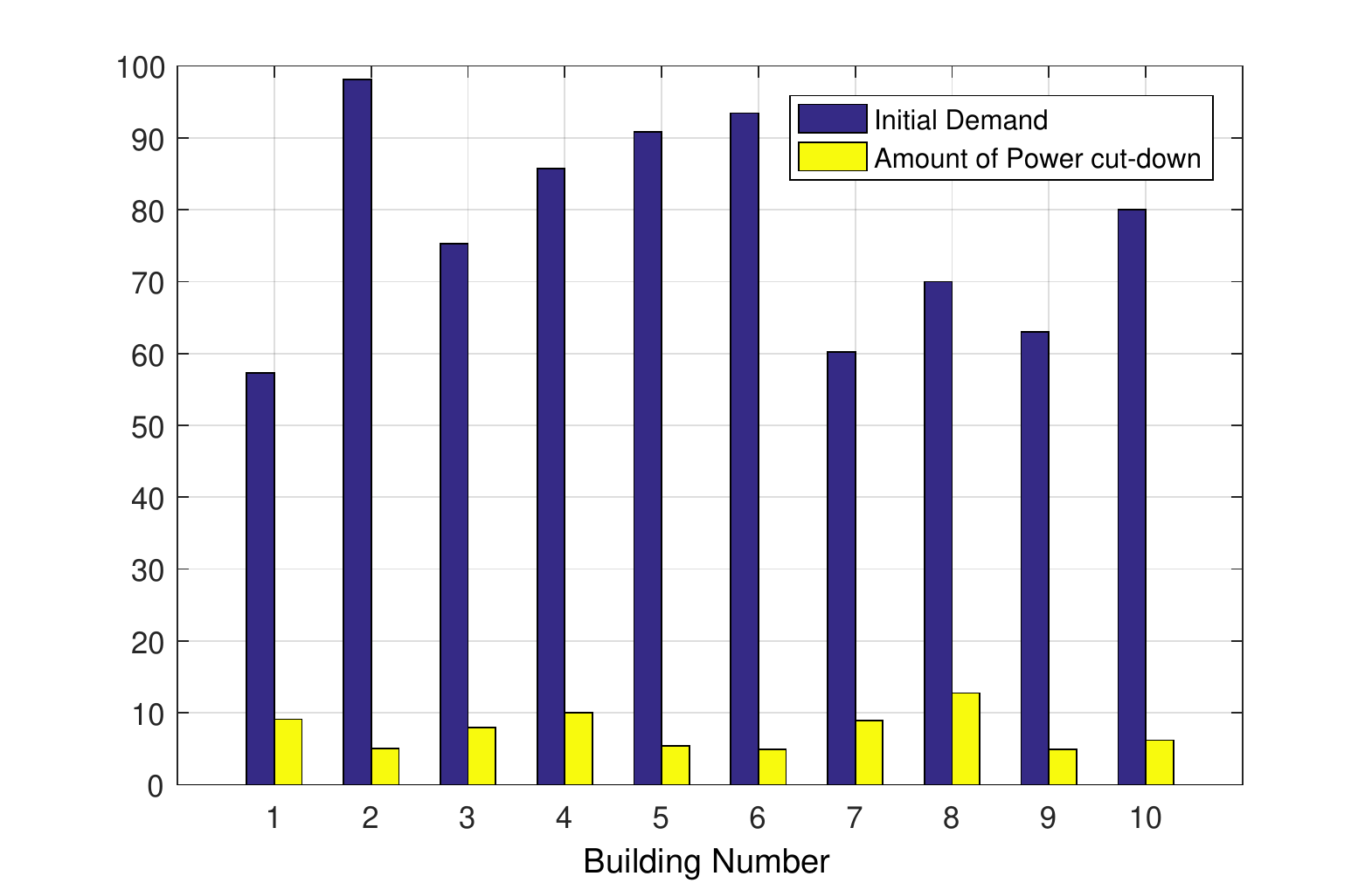}
	\caption{Demand cut down by the 10 buildings.}
	\label{fig:8}
	%	\hrule
\end{figure}

The static approach assumes that each building predicts its initial demand (for the next hour or minute, depending on the time granularity) at the beginning of a time window, which will be available for consumption at the end of that time window. Unlike the static approach, if the predicted demand of a building changes during a time window, it will be accommodated in the next time window in the dynamic approach. Simulation results of this approach, where the length of the time window is smaller than the static case, using the same parameters as in sub-section~\ref{subsec:sim_static}, is shown in Fig.~\ref{fig:5} to Fig.~\ref{fig:8}. The results obtained are similar to that in the static approach, but with the potential adapt to dynamic changes in the predicted demand or grid maximum.
%

%%%%%%%%%%%%%%%%%%%%%%%%%%%%%%%%%%%%%%%%%%
%%%%%%%%%%    Section VIII    Conclusions         %%%%%%%%%%%%%%%
%%%%%%%%%%%%%%%%%%%%%%%%%%%%%%%%%%%%%%%%%%
\section{Conclusions}
\label{sec:conclusion}
This work proposed a distributed model to continuously maintain a stable grid by co-operative action of all the buildings in the grid. Subject to the constraint that the total power consumption of all the buildings at any time is always less than the grid maximum, a pricing scheme has been formulated that will allow individual buildings to adapt to price estimates and adjust their demands to maximize their own benefit. Simulation results have demonstrated the convergence of the algorithm. From the results, we saw that when the system reaches equilibrium, the demands of the individual buildings reaches its optimal demand and the final aggregate demand is close to but less than grid maximum. Maintaining the aggregate demand just close to the grid maximum prevents under-utilization of the grid capacity.

%\section*{Acknowledgment}

% Bibliography Section
\bibliographystyle{IEEEtran}
\bibliography{references}

\begin{thebibliography}{1}
\providecommand{\url}[1]{#1}
\csname url@samestyle\endcsname
\providecommand{\newblock}{\relax}
\providecommand{\bibinfo}[2]{#2}
\providecommand{\BIBentrySTDinterwordspacing}{\spaceskip=0pt\relax}
\providecommand{\BIBentryALTinterwordstretchfactor}{4}
\providecommand{\BIBentryALTinterwordspacing}{\spaceskip=\fontdimen2\font plus
\BIBentryALTinterwordstretchfactor\fontdimen3\font minus
  \fontdimen4\font\relax}
\providecommand{\BIBforeignlanguage}[2]{{%
\expandafter\ifx\csname l@#1\endcsname\relax
\typeout{** WARNING: IEEEtran.bst: No hyphenation pattern has been}%
\typeout{** loaded for the language `#1'. Using the pattern for}%
\typeout{** the default language instead.}%
\else
\language=\csname l@#1\endcsname
\fi
#2}}
\providecommand{\BIBdecl}{\relax}
\BIBdecl

\bibitem{kelly1998rate}
F.~P. Kelly, A.~K. Maulloo, and D.~K. Tan, ``Rate control for communication
  networks: shadow prices, proportional fairness and stability,'' \emph{Journal
  of the Operational Research society}, vol.~49, no.~3, pp. 237--252, 1998.

\bibitem{ganesh2001congestion}
A.~Ganesh, K.~Laevens, and R.~Steinberg, ``Congestion pricing and user
  adaptation,'' in \emph{INFOCOM 2001. Twentieth Annual Joint Conference of the
  IEEE Computer and Communications Societies. Proceedings. IEEE}, vol.~2.\hskip
  1em plus 0.5em minus 0.4em\relax IEEE, 2001, pp. 959--965.

\bibitem{paschalidis2011market}
I.~C. Paschalidis, B.~Li, and M.~C. Caramanis, ``A market-based mechanism for
  providing demand-side regulation service reserves,'' in \emph{2011 50th IEEE
  Conference on Decision and Control and European Control Conference}.\hskip
  1em plus 0.5em minus 0.4em\relax Ieee, 2011, pp. 21--26.

\bibitem{weng2011managing}
T.~Weng, B.~Balaji, S.~Dutta, R.~Gupta, and Y.~Agarwal, ``Managing plug-loads
  for demand response within buildings,'' in \emph{Proceedings of the Third ACM
  Workshop on Embedded Sensing Systems for Energy-Efficiency in
  Buildings}.\hskip 1em plus 0.5em minus 0.4em\relax ACM, 2011, pp. 13--18.

\bibitem{fan2011distributed}
Z.~Fan, ``Distributed demand response and user adaptation in smart grids,'' in
  \emph{12th IFIP/IEEE International Symposium on Integrated Network Management
  (IM 2011) and Workshops}.\hskip 1em plus 0.5em minus 0.4em\relax IEEE, 2011,
  pp. 726--729.

\bibitem{xiao2004fast}
L.~Xiao and S.~Boyd, ``Fast linear iterations for distributed averaging,''
  \emph{Systems \& Control Letters}, vol.~53, no.~1, pp. 65--78, 2004.

\bibitem{rajagopalan2011distributed}
S.~Rajagopalan and D.~Shah, ``Distributed averaging in dynamic networks,''
  \emph{IEEE Journal of Selected Topics in Signal Processing}, vol.~5, no.~4,
  pp. 845--854, 2011.

\bibitem{nedic2009distributed}
A.~Nedic, A.~Olshevsky, A.~Ozdaglar, and J.~N. Tsitsiklis, ``On distributed
  averaging algorithms and quantization effects,'' \emph{IEEE Transactions on
  Automatic Control}, vol.~54, no.~11, pp. 2506--2517, 2009.

\end{thebibliography}

\end{document}